\RequirePackage{fix-cm}
\documentclass[smallcondensed]{svjour3}     
\smartqed  
\usepackage{amsmath,amssymb,amsfonts}
\usepackage{algorithmic}
\usepackage{graphicx}
\usepackage{textcomp}
\usepackage{xcolor}

\usepackage{algorithm}
\usepackage{subfigure}

\usepackage{diagbox}
\usepackage{threeparttable}
\usepackage{multirow}
\usepackage{color}
\usepackage{url}
\usepackage[misc]{ifsym}

\usepackage{booktabs}
\usepackage{caption}

\usepackage{setspace}
\usepackage{bbold}

\begin{document}

\title{Improving Specificity in Mammography Using Cross-correlation between Wavelet and Fourier Transform
}

\titlerunning{Improving Specificity in Mammography}        

\author{Liuhua Zhang
}

\authorrunning{L. Zhang} 


\institute{Liuhua Zhang \\
	{lz4038ca@gmail.com} 
}

\date{Received: date / Accepted: date}

\maketitle

\begin{abstract}

Breast cancer is in the most common malignant tumor in women. It accounted for 30\% of new malignant tumor cases. Although the incidence of breast cancer remains high around the world, the mortality rate has been continuously reduced. This is mainly due to recent developments in molecular biology technology and improved level of comprehensive diagnosis and standard treatment. Early detection by mammography is an integral part of that. The most common breast abnormalities that may indicate breast cancer are masses and calcifications. Previous detection approaches usually obtain relatively high sensitivity but unsatisfactory specificity. We will investigate an approach that applies the discrete wavelet transform and Fourier transform to parse the images and extracts statistical features that characterize an image’s content, such as the mean intensity and the skewness of the intensity. A na\"{\i}ve Bayesian classifier uses these features to classify the images. We expect to achieve an optimal high specificity.

\end{abstract}

\section{Introduction}

\subsection{Breast Cancer}
Breast cancer is the most commonly diagnosed form of cancer in women and the second-leading cause of cancer-related death behind lung cancer [2]. According to the American Cancer society, “over 200,000 new cases of invasive breast cancer are diagnosed each year, and there are nearly 40,000 women who died of breast cancer in 2011”. Basically, there are two types of breast cancer, including invasive and noninvasive. Invasive cancer has spread from the milk duct or lobule to other tissues in the breast. Noninvasive cancer has not yet invaded other breast tissue. It is called ``in situ.''

Breast cancer can be diagnosed at different stages of woman’s life and may expand at different rates. For example, breast cancer occurs from twenty years old in developed countries such as North Europe and North America. Its incidence trends to double every ten to twenty years, and it probably reaches a peak between 75 and 85 years old. Nevertheless, in less developed countries such as Asia, the highest peak of normal breast cancer ranges between 45 and 55. Whatever the country is, breast cancer incidence rate is constantly increasing. The earlier these cancers are discovered, the higher the chance of survival. This leads to the rational for using mammographic screening.

Early signs of breast cancer, including masses and calcifications, can be detected by screening mammography. Masses are regions of unusual tissue, and they can be benign (usually round or oval shape), or malignant (mostly spiculate shape). Breast calcifications are calcium deposits (usually look like white spots or flecks on a mammogram) that appear within the soft tissue of breast [26], and they can also be divided into two kinds: Macrocalcifications and Microcalcifications. Macrocalcifications are bigger clusters of calcium deposits, and are usually not associated with breast cancer. On the other hand, Microcalcifications are tiny clusters of calcium deposits which indicate the presence of breast cancer [5]. Four examples are shown below in figure 1(a), (b), (c) and (d).

\begin{figure}[!htbp]
	\centering
	\includegraphics[height=\textheight,width=\textwidth]{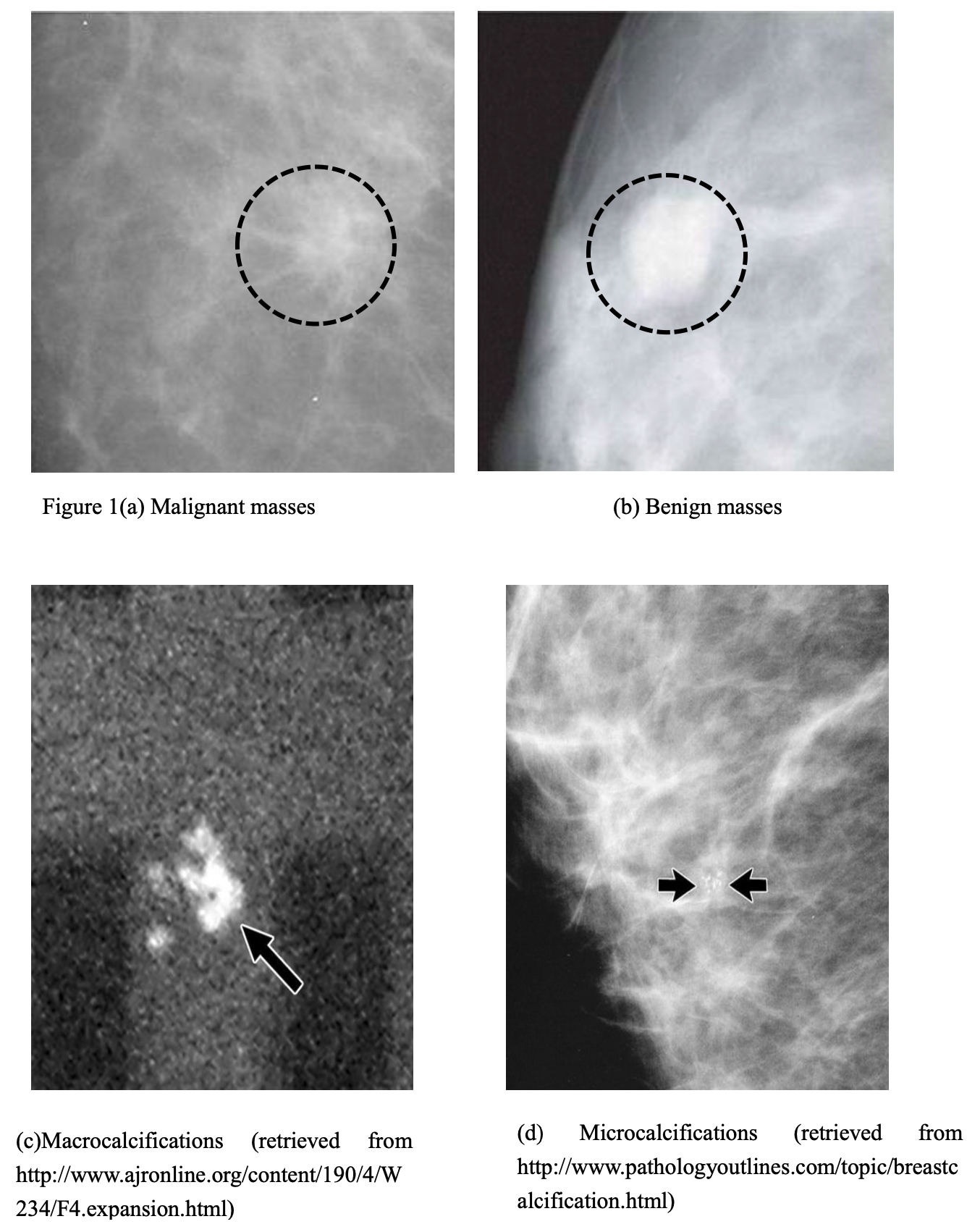}
	\caption{Digital mammograms illustrating four kinds of breast cancer. (a) malignant masses with spiculate shape (b) benign masses with round shape (c) macrocalcifications in bigger clusters of calcium deposits (d) microcalcifications in tiny clusters of calcium deposits.}
	\label{fig:fairCalibration}
\end{figure}

\subsection{Mammography}
Mammography is a ``specific type of imaging that uses a low-dose x-ray system to examine breast''[3]. To date, screening mammography is the best available radiological technique for early detection of breast cancer [1]. A mammography exam, called a mammogram, is used to aid in the early detection and diagnosis of breast diseases in women. Two recent advances in mammography include digital mammography and computer-aided detection. Digital mammography, also called full-field digital mammography (FFDM), ``is a mammography system in which the x-ray film is replaced by solid-state detectors that convert x-rays into electrical signals'' [27]. Digital mammography detects similar results with digital cameras. For that reason, it is essentially no difference whether a patient takes a digital mammogram or a conventional film mammogram. However, digital mammography is less time-consuming and may proceed with more mammograms in a day. It helps patients save time in waiting to see doctors and get their results in time.

Computer-aided detection (CAD) systems operate though the imaging, medical image processing technology and other possible physiological, biochemical methods, combining computer calculation and analysis. The purpose of CAD software is to assist doctors in detecting disease and improving the diagnostic accuracy. Specifically, a conventional film mammogram or a digital mammogram passes a mammographic image to the CAD system. It then searches for any abnormal areas such as density and calcification that may indicate the pathology of breast cancer. These suspicious areas on the images will be marked out by the CAD system, which could be a sign for the radiologist in further analysis.

Generally, a mammographic image could have two basic views: craniocaudal (CC) view which is taken from above a horizontally-compressed breast and mediolateral-oblique (MLO) view which is taken from the side and at an angle of a diagonally-compressed breast. They are shown in Figure 2(a) and Figure 2(b), respectively.

\begin{figure}[!htbp]
	\centering
	\includegraphics[height=0.3\textheight,width=\textwidth]{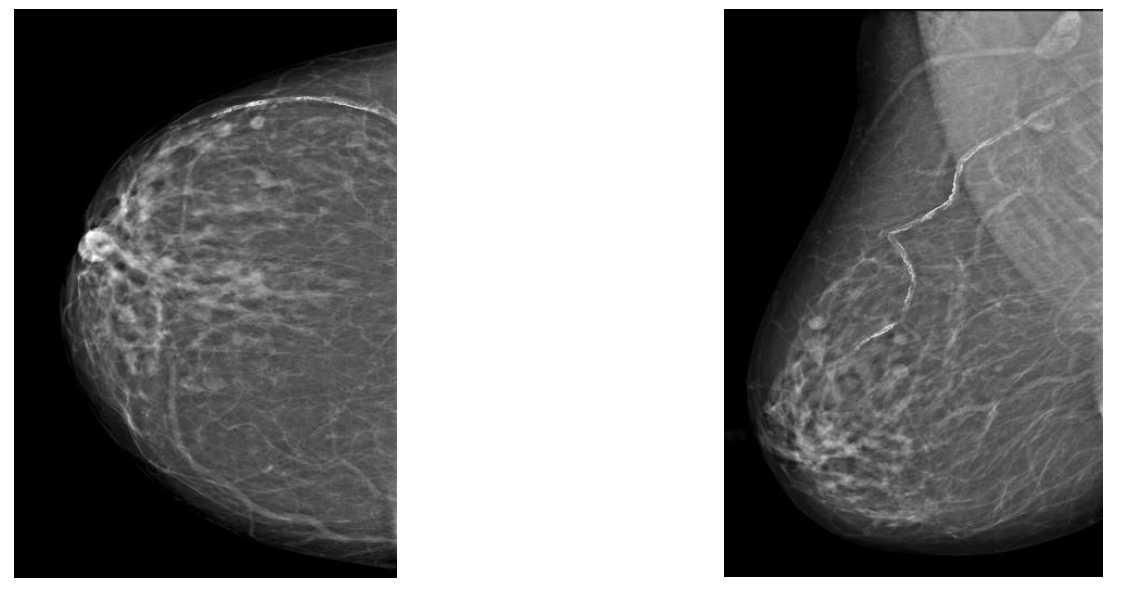}
	\caption{Digital mammograms illustrating the conventional views of the breast. a) Craniocaudal view (CC) the compressed breast is viewed from above; (b)mediolateral oblique (MLO) the compressed breast is viewed laterally towards the midline. (retrieved from one case in the DDSM database)}
	\label{fig:fairCalibration}
\end{figure}

Mammography plays a central role in early detection of breast cancers because it can show changes in the breast up one or two years before a patient or physician can feel them [27]. Current guidelines from the U.S. Department of Health and Human Services (HHS), the American Cancer Society (ACS), the American Medical Association (AMA) and the American College of Radiology (ACR) recommend women older than 40 take screening mammography every year. It is reported that ``annual mammograms'' help to make breast cancer easier to be detected and cured if the therapies are effective.

\subsection{Terminology of diagnosis rates}
The performance of such a mammography screening system can be measured by two parameters: sensitivity and specificity.

Sensitivity can be also called the true positive rate; it is the proportion of the cases being abnormal when breast cancer is present. For example, if 100 women did have breast cancer among 1000 screened patients, and a mammography screening tool detected cancer in 90 cases, then the sensitivity is 90/100 or 90\%. Sensitivity may
depend on several factors, such as lesion size, breast tissue density, or overall image quality. In cancer screening protocols, sensitivity is deemed more important than specificity, because failure to diagnose breast cancer may result in serious health consequences for a patient. Almost fifty percent of cases in medical malpractice relate to ``false-negative mammograms'' [4].

Specificity is also known as the true negative fraction; it is the proportion of the cases being normal when breast cancer is absent. For example, if 100 cases of breast cancer were diagnosed in a set of 1000 patients, and the screening system finds 720 cases to be normal and 180 uncertain in the remaining 900 normal patients, it can be calculated that the specificity of the mammography tool is 720/900 or 80\%. Although the consequences of a false positive, that is, diagnosing a normal patient as having breast cancer, are less severe than missing a positive diagnosis for cancer, specificity should also be as high as possible. False positive examinations can result in unnecessary follow-up examinations and procedures and may lead to significant anxiety and concern for the patient.

\section{Research Actuality}

\subsection{Feature Extraction}
A feature is an individual measurable heuristic property that can characterize a specific region or pathology. In general, features can be divided into three categories: intensity features, geometric features and texture features.

Wavelet methods have been researched since late 1990s. In 1996, Chen, C.H. and Lee, G.G. [6] implemented a multi-resolution orthogonal basis wavelet scheme to extract effective features. Those selected features were used for classification through back propagation neural network (BPNN) and fuzzy c-means (FCM) classification. This experiment showed encouraging results. Recently, a new research developed a novel approach which exploits the 4-level Daubechies 4 wavelet features of ``radial and circular scan lines drawn over the region of interest (ROI)'' [7]. Then they were tested in a mini Mammography Image Analysis Society (MIAS) database using three
different kinds of classifiers, including Support Vector Machine (SVM), two layer feed-forward Neural Network, and Bayesian. Results showed SVM classifier achieved the highest accuracy rate (85.96\%).

Similarly, Fourier transform has always been studied in feature extraction and compared with wavelet methods. The Fourier transform maps a signal from the time domain into the frequency domain, whereas a wavelet transform maps a signal into a two dimensional time-frequency domain. This different kind of mapping provides information about a signal’s frequency content as a function of time, in contrast to the time-independent output of the Fourier transform. Recent research has shown that a novel group of features extracted by Fast Fourier Transform (FFT) based on the Radial Distance (RD) achieved promising results. In the experiment, 40 frequently appeared features were selected and trained in ``Multilayer Perception (MLP) network with Back Propagation learning rule''. Performances were evaluated by the ROC curves, and this schema obtained an area under the ROC curves (Az) which was equal to 0.98 [8].

Besides intensity features, boundary features can also be used in mammography. Boujelben, A., Chaabani, A.C., Tmar, H., and Abid, M. analysed contour information, including Radial Distance, Index Angle, and convexity. Two techniques of classification (k-Nearest Neighbors and MLP) were used to classify mass region. 94.2\% sensitivity and 97.9\% specificity were obtained by kNN classifier and MLP classifier separately [9].

\subsection{ Feature Selection}
Selecting the significant features is rather essential for breast cancer classification. A feature is regarded as significant based on its discrimination, reliability, independence and optimality [10]. Several feature selection methods have been developed and employed in mammography CAD systems, and they can be divided into three categories: filter methods, wrapper methods, and hybrid methods. Filter methods search for significant features based on an independent test; while wrapper methods apply a specific machine learning algorithm. The so called wrapper
refers to the use of the existing supervised learning classifier (decision tree, neural networks, SVM and so on) so that it can effectively solve the learning situation that only a small amount of marked samples exist in lots of no mark samples. Hybrid methods search for features by employing both an independent test and a performance evaluation function.

Pethalakshmi, A., Thangavel, K., and Jaganathan K. [11] applied rough set theory based reduction algorithms such as Quickreduct (QR), and proposed Modified Quickreduct(MQR) in reducing the mammography features. The algorithms were measured in MIAS database and proved to be successful and efficient in containing important information.

Peng, Z., Verma, B., and Kumar, K. [12] developed a genetic algorithm for feature selection, where experiments were taken through neural network. Different feature subsets were applied to calcification and mass cases, and they obtained 90.5\% and 81.2\% accuracy rate respectively.

In 2011, Yihua Lan, Haozheng Ren, Yong Zhang, Hongbo Yu, and Xuefeng Zhao [13] proposed a hybrid method which firstly used a filter method, linear discriminative analysis (SLDA) algorithm, to remove irrelevant features, and then used a wrapper-based approach, genetic algorithm (GA) to remove useless features and achieve the ultimate feature subset. Compared with single SLDA or GA selection method, the proposed one could achieve better classification accuracy result with not even one half features.

\subsection{Classification Methods}
After a set of features are selected, the last challenging task is automatic classification of masses or microcalcifications. Classification is vital because it relates to the final result, that is, whether a patient is normal or cancerous. Many classifiers have been developed for classification, among which the most commonly used classification methods are: neural networks, Bayesian classification, K-nearest neighbor classifiers, support vector machine and different decision trees.

In 2005, Gholamali Rezai-rad [14] developed an intelligent system based on wavelet and neural network for the identification of microcalcification clusters in digitized mammograms. The system was tested in the Nijmegen and the MIAS Mammographic databases with satisfactory results. The achieved classification specificity was 1.76 and 1.12 false positive clusters per image, for the Nijmegen and MIAS dataset, respectively, at the sensitivity level of about 0.94.

Maurice Samulski, Nico Karssemeijer, Peter Lucas, and Perry Groot [15] compared support vector machines (SVMs) and Bayesian networks (BNs). Bayesian classifier achieved significant improvement in accuracy after applying principal component analysis (PCA) for reducing high-dimensional data, while SVM classifier had no obvious increase. Both of the classifiers performed similar accuracy result after transformation.

In [16], the authors presented a SVM based computer-aided diagnosis (CAD) system which features based on shape, texture and statistical properties are extracted from each region for the characterization of clustered microcalcifications in digitized mammograms. Finally, these features are fed to a SVM based classifier for identifying the clusters as either benign or malignant. The SVM with RBF kernel gave Az = 0.9803 with 97\% accuracy and the SVM with polynomial kernel gave Az = 0.9541 with 95\% accuracy.

\section{Research questions}
According to the related literature review, two commonly used methods in selecting mammograms features are wavelet transform and Fourier transform. So far, both methods can achieve high sensitivity ($>$98\%) in cooperating with most classifiers. However, the result of specificity is not that satisfactory. It fluctuates from 60\% to 80\%. Sometimes a proposed system showed acceptable sensitivity and specificity, but the computation cost a lot of time. And among those experiments, seldom approaches use both wavelet transform and Fourier transform together. This leads to the fundamental research question:

\textbf{Will cross-correlation improve the specificity of the classification regime?}

In this classification regime, we will experiment the Na\"{\i}ve Bayesian classifier first to see if it is effective. With regard to the statistical features we use in the process of classification, we will combine the wavelet transform and Fourier transforms together to extract the mean intensity, the standard deviation of the pixel intensities, the skewness of the pixel intensities and the kurtosis of the pixel intensities. After we obtain all eight features, we then select the most fitting features to do the classification task.

\section{Research objectives and method}

\subsection{Research objectives}
The primary objective of this research is to design a tool that combine two kinds of wavelet transform together in selecting optimal features and improve the final specificity rate. Specific research objectives are:

\begin{itemize}
	\item[1.] Develop a set of pre-processing steps to isolate the tissue in the images and regularize the appearance of the images to make direct comparisons possible.
	\item[2.] Apply the wavelet transform and Fourier transform to parse an image and generate a set of scalar features based on the output of the transform to characterize each image.
	\item[3.] Classify the images as normal or suspicious and give the specificity of the result.
	\item[4.] Evaluate and compare the performance with previous studies.
\end{itemize}

\subsection{Database}
Data was obtained from two publicly available mammographic image databases. The digital database for screening mammography (DDSM) [18][19] and the mammographic image analysis society (MIAS) database [17]. These two databases are extremely useful for development and testing of computer aided detection schemes. The two databases contain images saved in different formats, with different resolutions, collected on different machines, by different technicians. The MIAS database consists of 322 mediolateral oblique images (161 left, right pairs) with ground truth data for each one. Images are characterized according to density, class of abnormality, and the severity of the abnormality present. The DDSM database which contains 2620 cases is much larger than the MIAS. The cases are partitioned into 43 volumes. 12 normal, 15 cancer, 14 benign, and 2 benign without callback. All images used in this research are stored in PGM format. The DDSM has three main advantages over the MIAS database. It has more images, it has higher resolution images, and it has craniocaudal images in addition to mediolateral oblique. Its only downside is that it is not nearly as user friendly.

\subsection{Image analysis system}
The entire image analysis system, from the reading of the original image to the final classification as either normal or suspicious consists of two distinct stages. The first stage is the image processing system, which reads in the original image and produces a set of wavelet map images for the classifier to use. The second stage is the classification system, which measures features from the wavelet images and classifies the image as either normal or suspicious based on the results from the ensemble of classifiers. It can be represented by the block diagram of the following Figure 3:

\begin{figure}[!htbp]
	\centering
	\includegraphics[height=0.2\textheight,width=\textwidth]{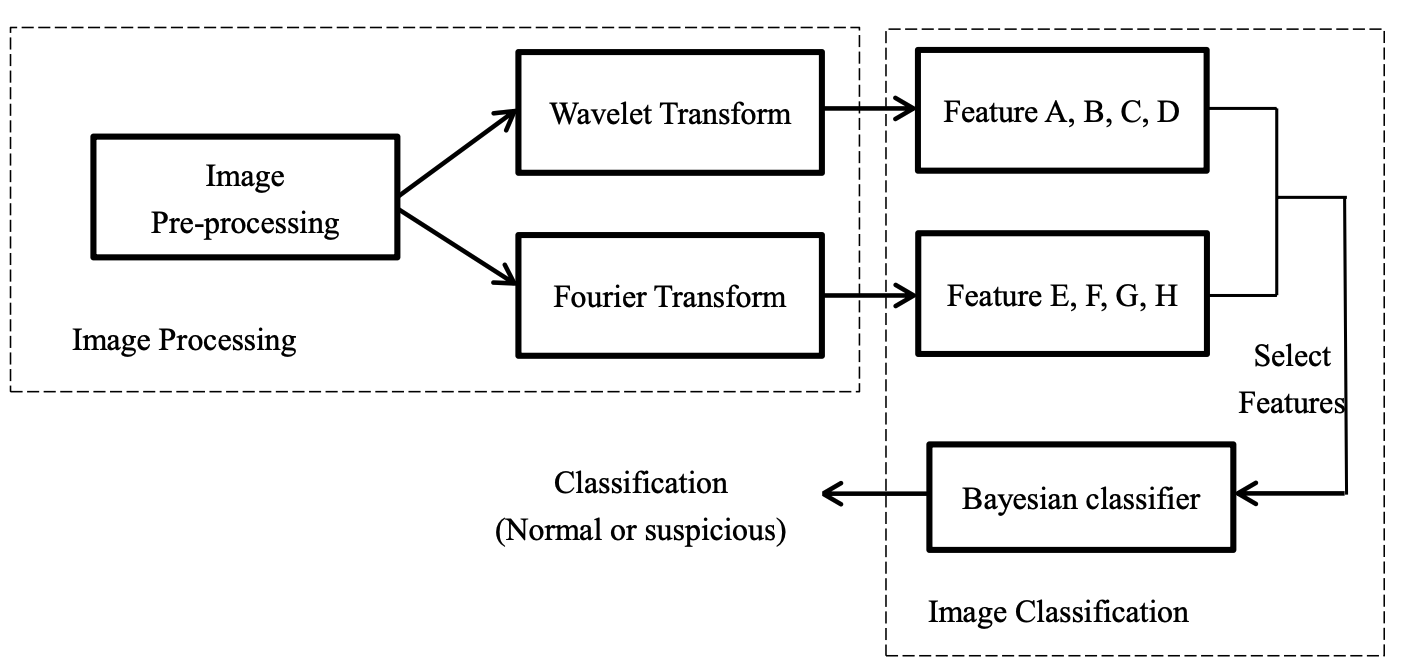}
	\caption{Image analysis system (The analysis consists of two major sub-systems and each of these houses three distinct functions).}
	\label{fig:fairCalibration}
\end{figure}

\subsubsection{Image Pre-processing}

In order to reduce the influence of information content not related to pathology, several pre-processing steps are implemented to regularize the appearance of the images and remove any unnecessary artefacts. The steps taken were: orientation matching, background thresholding, artefact removal and intensity matching.

Orientation matching ensures that all images pointed in the same direction, preventing changes in the wavelet transform coefficients due only to the directionality change between right and left images. The thresholding process is implemented in conjunction with the next step, artefact removal; because of this, it is convenient to perform a binary thresholding, where all pixels below the threshold are set to an intensity of zero and all pixels above the threshold are set to an intensity of one. After that, artefacts in the image can be removed easily. The last step, intensity matching, scales all images so that the intensity of the brightest pixel in the image has a relative intensity of 1.0 and linearly scales all other image pixels accordingly. This step ensures uniformity across different images, which may be taken at different times under slightly different machine settings or by different personnel.

\subsubsection{Wavelet Transform}

Wavelets are functions generated from one basis function called mother wavelet by scaling and translating in frequency domain. If the mother wavelet is denoted by $\psi(t)$, the other wavelets $\psi_{a,b}(t)$ can be represented as:

\begin{equation}
	\psi_{a,b}(t)= \frac{1}{\sqrt{|a|}}\psi(\frac{t-b}{a})
\end{equation} 

\noindent where a and b are two arbitrary real numbers. The discrete wavelets can be represented by: 

\begin{equation}
	\psi_{m,n}(t)= a_0^{-\frac{m}{2}}\psi(a_0^{-m}t- n_0b_0)
\end{equation} 

\noindent There are many choices to select the values of $a_0$ and $b_0$. Most commonly used values are $a_0$ = 2 and $b_0$ = 1.

\begin{figure}[!htbp]
	\centering
	\includegraphics[height=0.2\textheight,width=\textwidth]{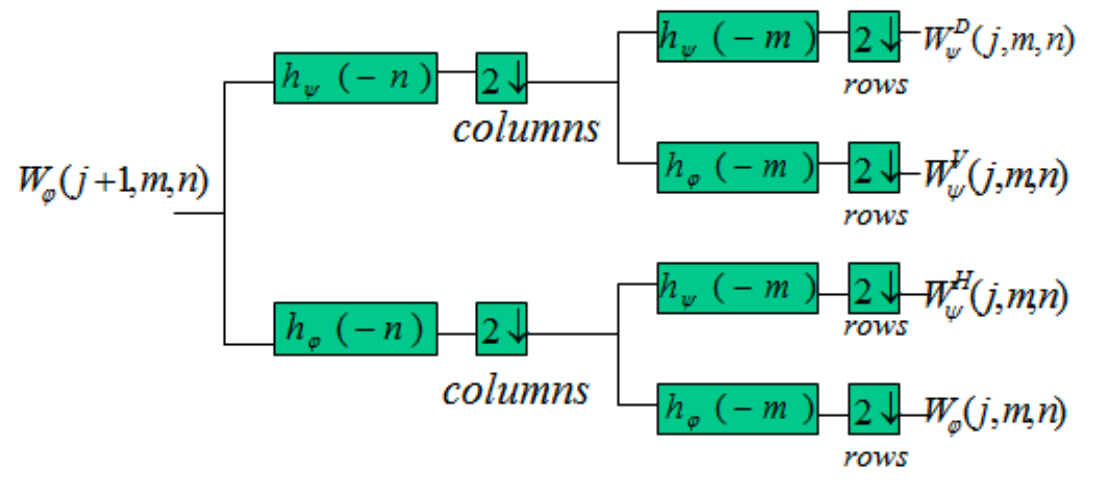}
	\caption{Fast 2D wavelet transform.}
	\label{fig:fairCalibration}
\end{figure}

Figure 4 illustrates the general form of 2D wavelet transform. The decompositions firstly run along the x-axis and then calculate along the y-axis, a picture then can be divided into four bands: LL (left-top), HL (right-top), LH (left-bottom) and HH (right-bottom). 

\subsubsection{Fourier Transform}

The Fourier Transform is an important image processing tool which is used to decompose an image into its sine and cosine components. The output of the transformation represents the image in the Fourier or frequency domain, while the input image is the spatial domain equivalent. In the Fourier domain image, each point represents a particular frequency contained in the spatial domain image.

The Discrete Fourier Transform (DFT) is the sampled Fourier Transform and therefore does not contain all frequencies forming an image, but only a set of samples which is large enough to fully describe the spatial domain image. The number of frequencies corresponds to the number of pixels in the spatial domain image, i.e. the images in the spatial and Fourier domain are of the same size.

For a square image of size N×N, the two-dimensional DFT is given by:

\begin{equation}
	F(k,l)= \sum_{N-1}^{i=0}\sum_{N-1}^{j=0} f(i,j) e^{-i2\pi(\frac{ki}{N}+\frac{lj}{N})}
\end{equation} 

\noindent where $f(a,b)$ is the image in the spatial domain and the exponential term is the basis function corresponding to each point $F(k,l)$ in the Fourier space. The equation can be interpreted as: the value of each point $F(k,l)$ is obtained by multiplying the spatial image with the corresponding base function and summing the result.

\subsubsection{Features}

In this experiment, we will extract four statistical features: the mean intensity, the standard deviation of the pixel intensities, the skewness of the pixel intensities and the kurtosis of the pixel intensities. Further, the classification system will use some of these features to classify whole images as being normal or suspicious.

\begin{itemize}
	\item[1.]  Mean. The mean, m of the pixel values in the defined window, estimates the value in the image in which central clustering occurs. 
	\item[2.]  Standard Deviation. The Standard Deviation is the estimate of the mean square deviation of grey pixel value $I(i,j)$ from its mean value. Standard deviation describes the dispersion within a local region.
	\item[3.] Skewness. The third statistic measured from each wavelet map image is the skewness of the pixel intensities. The skewness of a distribution of values is defined as the third central moment of the distribution, normalized by the cube of the standard deviation.
	\item[4.] Kurtosis. The fourth and final statistic measured from the wavelet maps is the kurtosis of the pixel intensities. The kurtosis of a distribution of values is defined as the fourth central moment of the distribution, normalized by the fourth power of the standard deviation of the distribution
\end{itemize}

\subsubsection{Na\"{\i}ve Bayesian classifier}

In numerous classification models, the most widely used one is Naive Bayesian classifier (NBC). Naive Bayesian model, which originated from classical mathematics theory, has a solid foundation of mathematics and stable classification efficiency. At the same time, NBC model has a simple algorithm which needs a few estimated parameters. It is not too sensitive with the missing data. In theory, NBC model has a smaller error rate compared with other classification methods. More specifically, NBC model performs the best when the attribute values are independent.

First, the Bayes theorem is the foundation.

\begin{equation}
	P(B|A)= \frac{P(A|B)P(B)}{P(A)}
\end{equation} 

\noindent Here, $P(A|B)$ is called conditional probability, which refers to the probability that A happens when B has already happened.

The idea of Naive Bayesian classifier is: For the undetermined classification class, calculate probabilities that the class belongs to each category. It is then classified to the class which has the largest probability. In common words, when you see a black man in street, most people will guess he may come from Africa. Because Africa has the highest ratio in black, the category with the largest conditional probability will be chosen without other available information.

 \subsection{Evaluation}
 
 The entire evaluation process contains two parts:

 \begin{itemize}
 	\item[1.] The specificity of mammography is the main purpose of my thesis. In this binary classification problem, a classifier yields two discrete results: positive and negative. As shown in Figure 6, there could be four outcomes for a classifier in judging a sample. We named the four results in confusion matrix as true positive (TP), true negative (TN), true positive (TP), false negative (FN). ``TP'' means a positive instance is classified correctly as positive; ``FN'' refers to the positive instance wrongly classified as negative. Similarly, ``TN'' implies a negative instance is correctly classified as negative; otherwise it is ``FP''.

 	\item[2.] To strengthen the confidence of this classification method and compare it with the classifiers in previous studies, we plan to further estimate under the Receiver Operating Characteristics (ROC) curves. ROC illustrates the performance of a binary classifier system and it is created by sensitivity vs. specificity. To date, it has achieved excellent performances in the related fields especially in medicine and radio [20–23].
 \end{itemize}

A good classifier would produce an ROC curve which locates closer to the upper left hand corner. In other words, the space left under a perfect classifier curve is always large. To simplify the measurement of how well a classifier implements, the area under the curve (AUC) is used [24, 25]. For example, the classifier `a' always outperforms the classifier `b' as shown in Figure 7 according to the definition.

In general, the area under the ROC curve ranges between 0.5 and 1.0, and the closer AUC equals to 1, the better the classifier is in this condition. The classifier
achieves lower accuracy when AUC is between 0.5 and 0.7; it has certain accuracy when AUC is in 0.7$\sim $0.9, and higher accuracy if AUC is above 0.9. The classifier completely does not work if AUC equals to 0.5. The situation that AUC is lower than 0.5 seldom appears in real life.

\section{Conclusion}

This work is in the exploration of the benefit of combining two kinds of wavelet transform together. We will build a novel mammographic image analysis system integrating wavelet and Fourier transform and measure the benefits of the approach. We expect that the proposed system delivers better performance than the previous studies.


%
%




\end{document}